# How to increase the physics output per MW.h for FCC-ee?
## Parameter optimization for maximum luminosity

D. Shatilov[1,2,a]

[1] BINP, Novosibirsk, Russia
[2] CERN, Geneva, Switzerland

**Abstract** The efficiency of colliders for physics is largely determined by their luminosity, while most of the energy consumed by high-energy $e^+e^-$ colliders is proportional to the total beam current. Thus, the energy efficiency is mainly determined by the specific luminosity that needs to be maximized. One of the most effective ways to achieve this is by using the Crab waist collision scheme, which implies a large Piwinski angle (LPA). A distinctive feature of the FCC-ee is the great influence of beamstrahlung (radiation in the field of an opposite bunch) on beam dynamics. At low energies, this manifests itself in a significant increase in the energy spread and bunch length, at high energies, in a limitation of the beam lifetime. The collision of intense bunches with LPA and beamstrahlung can also lead to various kinds of instabilities limiting the luminosity. Here, we discuss the main aspects to consider when optimizing the parameters of the FCC-ee collider at different energies and explain the choice of basic parameters such as RF voltage, lattice functions at IP, bunch intensity, etc. We will also pay attention to open issues requiring further study and identify some key points for the next phase of this project.

## 1 Collider size, beam energy and total beam current

The goals and objectives for the FCC-ee are determined by the open questions in particle physics that also informed the recent update of the European Strategy. The proposed physics programme of FCC-ee requests operating in the center-of-mass energy range from 91 to 365 GeV. This is only the first step of the integrated FCC project, foreseen an energy-frontier hadron collider (FCC-hh) housed in the same tunnel. It is designed for achieving the maximum possible energy, which is proportional to the radius of the ring. Terrain is one of the main limiting factors here, which is why the FCC is as large as possible to fit between the mountains around CERN. But even if the ring size is already set, it is useful to see if it is optimal for the $e^+e^-$ collider. Here, the maximum energy is also determined by the average bending radius in the arcs $\rho$. However, in that case the limitation is not the magnetic field of the dipoles, but the energy loss per revolution due to synchrotron radiation (SR), which is proportional to $\gamma^4/\rho$. For example, in the LEP/LHC tunnel at $t\bar{t}$ energy it will amount to more than 35 GeV. To compensate for this energy loss, the RF cavities will have to be placed in numerous sections, which will occupy a noticeable part of the tunnel, thereby reducing the relative length occupied by the dipoles (filling factor). As a result, the bending radius in dipoles will decrease even more, thus increasing the SR losses and the length of the RF sections. Obviously, this sets a certain limit, and for a noticeable increase in the beam energy, it is necessary to increase $\rho$ and hence the size of the collider. There are many more reasons in favor of this but we will name only two:

– Reduction in the energy of emitted SR photons, which is necessary to keep the radiation background within the required limits.
– Reducing emittances and energy spread in the beams. The first is important for luminosity, the second is for accurate energy measurement by resonant depolarization. For example, the mass of the W-boson could not be measured by this method at the LEP [1], but this will become possible at the FCC-ee precisely due to the increase in the collider size.

Finally, consider the effect of size on luminosity and energy efficiency, which is the ratio of luminosity to total power consumption. We note straight away that the luminosity and SR energy losses are proportional to the total beam current $I_{tot}$. First, let us estimate the dependence of the energy consumption on the ring size *at fixed values* of the beam energy (determined by the experiment) and $I_{tot}$. The main components of the consumed energy: (i) RF power —replenishment of SR energy losses. For a fixed number of particles, it falls as a square of $\rho$, while the beam current falls linearly, so the power at a given beam current scales as $1/\rho$. (ii) The power consumption for cryogenics depends on the number of RF cavities, that is, the energy loss per turn. To a first approximation, it is also proportional to $1/\rho$. (iii) Power supply of dipoles. Their number is proportional, and the current (magnetic field) is inversely proportional to $\rho$. Since the power is quadratically dependent on the current, the energy consumption for the dipoles scales as $1/\rho$. (iv) Power supply of other elements of the magnetic system (quadrupoles, sextupoles, etc.). Assuming that the lattice is tuned to minimize the emittance, their number is, on average, proportional to the perimeter. Since the supply current does not depend on $\rho$, energy consumption is mainly determined by the number of elements and grows linearly with the perimeter. (v) Engineering infrastructure (cooling, ventilation, general services, etc.). As a first approximation, the power consumed by it is proportional to the perimeter. In fact, the dependence will be weaker, since the expenses for cooling per unit length decrease with increasing $\rho$. (vi) Injector complex (without booster). Since the number of particles in beams at a given $I_{\text{tot}}$ increases linearly with the perimeter, some part of the energy consumption will also be proportional to it. And the other part (e.g., power supply of magnetic elements) does not depend on the size of the collider. (vii) Booster that accelerates the beams to the energy of the experiment and therefore is located in the same tunnel as the collider. The first five points are also relevant for it. (viii) Detectors with their own infrastructure. Their energy consumption does not depend on the size of the collider.

---

[a] e-mail: dnshatilov@gmail.com (corresponding author)

**Table 1** Energy consumption in MW for accelerator subsystems at different energies

| Beam energy [GeV] | 45.6 | 80 | 120 | 182.5 |
| --- | --- | --- | --- | --- |
| Subsystem | Z | WW | ZH | $t\bar{t}$ |
| Cryogenics | 1 | 9 | 14 | 46 |
| RF | 163 | 163 | 145 | 145 |
| Magnets | 4 | 12 | 26 | 60 |
| Other | 91 | 93 | 97 | 103 |
| Total | 259 | 277 | 282 | 354 |

The optimized size of a circular lepton collider is determined by the ratio of the first three points to the next three and depends on the energy of the experiment, as well as on the limit on the allowable energy consumption. For the FCC-ee, the SR power is limited to 50 MW per beam (100 MW in total) at all energies [2, 283]. In addition, one must consider the efficiency of the power sources that convert the electrical energy into RF energy. It depends on the type of cavities, which will be different at low and high beam energies. For reference, Table 1 shows some of the energy consumption figures for the FCC-ee [2, 524].

Since the luminosity grows linearly with $I_{\text{tot}}$, and a noticeable share of energy consumption goes to needs that are not related to it, the most energy-efficient operation will be with the maximum allowable $I_{\text{tot}}$, that is, with the maximum luminosity. And the amount of SR losses determines whether a large ring size will be effective. For instance, a new collider for the production of Z, W and Higgs bosons can be built in the LEP/LHC tunnel. It will be much cheaper than the FCC-ee, and energy consumption excluding RF will also be lower. With a small $I_{\text{tot}}$, when the share of RF costs is small, its energy efficiency will also be small, but still higher than in the FCC-ee at the same $I_{\text{tot}}$. Yet our situation is different: high luminosity and energy efficiency require a large $I_{\text{tot}}$. Accordingly, more than 50% of the power (more than 60% at Z and W) will go to RF. But in the LEP/LHC tunnel, in order to stay within acceptable power limits, $I_{\text{tot}}$ (and hence the luminosity) must be several times smaller. It is easy to understand that energy efficiency in this case will be much lower. We also note that the luminosity and energy efficiency of FCC-ee are significantly higher than those of linear colliders in the same energy range [3,4].

## 2 Luminosity and collision scheme

In a symmetric collider (equal energies of both rings), the maximum luminosity is achieved with an equal population of colliding bunches. As will be shown later, this requirement is especially important for the FCC-ee. Therefore, in what follows we will assume that all bunches have the same number of particles, which means that their beam–beam parameters are also the same. In this case, the luminosity per IP for flat beams ($\sigma_y \ll \sigma_x$, both values here and below are the beam sizes at the IP) can be written as [5]

$$L = \frac{\gamma}{2er_e} \cdot \frac{I_{\text{tot}} \xi_y}{\beta_y^*} \cdot R_{hg} \qquad (1)$$

where $\xi_y$ is the vertical beam–beam parameter, $\beta_y^*$ is the vertical beta-function at the IP and $R_{hg}$ is the hour-glass factor. Of course, if the number of bunches is fixed, then the luminosity grows quadratically with the beam current. But it is more efficient to always have such a population of particles in a bunch that provides the maximum $\xi_y$. In this case, $I_{\text{tot}}$ will determine only the number of bunches, the luminosity linearly depends on $I_{\text{tot}}$, and it is convenient to redefine the specific luminosity as $L/I_{\text{tot}}$. The task of maximizing the luminosity is a priority for any collider. From the very beginning of their history, it was realized that one of the main limiting factors is the beam–beam interaction. Progress in collider performance was largely determined by how to increase the beam–beam limit (i.e., $\xi_y$), and how to get the maximum luminosity at a given $\xi_y$. Here, three main stages can be distinguished:

1. Decrease in $\beta_y^*$. In a head-on collision, this will require a corresponding decrease in the bunch length $\sigma_z$: otherwise $R_{hg}$ will become small and $\xi_y$ will have to be reduced. On the other hand, a decrease in $\sigma_z$ imposes a limitation on the number of particles in a bunch due to problems associated with impedance and collective instabilities.

**Fig. 1** Sketch of collision with large Piwinski angle

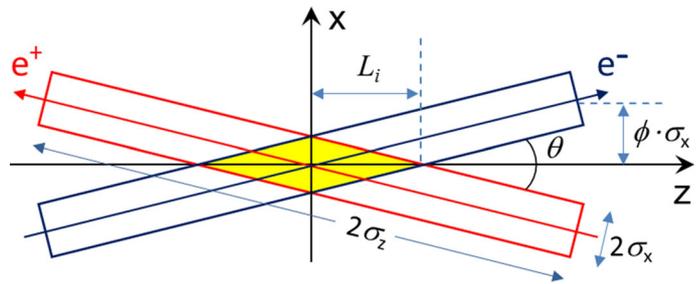

2. Two-ring colliders (factories) with a large number of bunches. This made it possible to increase $I_{tot}$ many times, but added its own problems: collision with a crossing angle, which leads to the excitation of synchro-betatron resonances, and long-range beam–beam interaction; both effects can limit $\xi_y$.
3. Crab waist (CW) collision scheme [6], which makes it possible to get rid of long-range beam–beam interaction, significantly reduce $\beta_y^*$ without enhancing the hour-glass effect, and raise $\xi_y$ several times.

The main parameter describing a collision with crossing angle $\theta$ is the Piwinski angle:

$$\phi = \frac{\sigma_z}{\sigma_x} tg\left(\frac{\theta}{2}\right) \tag{2}$$

A scheme of collision with LPA required for CW is shown in Fig. 1. The overlapping area of colliding beams (interaction area) is colored yellow and has an effective length $L_i \ll \sigma_z$, this is what allows us to get $\beta_y^* \ll \sigma_z$. And special "crab sextupoles" located on both sides of the IP suppress betatron coupling resonances and their synchrotron satellites [7], which allows to increase $\xi_y$. However, this requires a different approach to the selection of basic parameters. Indeed, $\xi_y$ for flat beams is given by [5]

$$\xi_y = \frac{r_e}{2\pi\gamma} \cdot \frac{N_p}{\sigma_x} \sqrt{\frac{\beta_y^*}{\varepsilon_y(1+\phi^2)}} \xrightarrow[\theta\ll 1,\ \phi\gg 1]{} \frac{r_e}{\pi\gamma\theta} \cdot \frac{N_p}{\sigma_z} \sqrt{\frac{\beta_y^*}{\varepsilon_y}} \tag{3}$$

Compared to the previous generation of colliders, we need at a "standard" bunch population $N_p$ to make $\phi \gg 1$, reduce $\beta_y^*$ (by about $\phi$ times), and increase $\xi_y$ several-fold. Obviously, this is possible only with small transverse beam sizes. As seen from the right-hand side of (3), $\xi_y$ (and hence the luminosity) is proportional to the linear charge density $N_p/\sigma_z$. This is the main parameter associated with impedance and collective instabilities, so it should not be too high. Accordingly, increasing $\xi_y$ is possible mainly due to a decrease in $\varepsilon_y$. The horizontal emittance is not directly involved, but since $\varepsilon_y \propto \varepsilon_x$, small $\varepsilon_x$ [2, 334] and small betatron coupling are required—similar to modern SR light sources.

## 3 Beamstrahlung

Small $\sigma_x$ required for CW collision greatly enhances beamstrahlung (BS), which can become crucial for high-energy colliders. The critical energy of BS photons is proportional to the strength of electromagnetic field from the counter bunch, which is proportional to its surface density:

$$u_c \propto \frac{\gamma^2 N_p}{\sigma_x \sigma_z} \tag{4}$$

Compared to older colliders, linear charge density has not changed much, but $\sigma_x$ makes the difference. For this reason, BS will be much stronger in FCC-ee than in LEP at the same energies. There are two effects here: a decrease in the beam lifetime due to single high-energy BS photons, and an increase in the energy spread (and hence the bunch length). The latter effect is enhanced with increasing $\rho$, since this decreases the contribution to the energy spread from SR in the arcs. And this is the second reason why the bunch lengthening due to BS in the FCC-ee will be so significant. The consequence of this may be a decrease in luminosity. For a given $\sigma_x$, BS depends on the linear charge density in the same way as $\xi_y$, and itself affects it through $\sigma_z$. To find an equilibrium steady state, we need to solve a self-consistent problem. For example, to get the "design" $\xi_y$ at low energy, $N_p$ should be 3.5 times larger than it would be without BS. Accordingly, $\sigma_z$ will grow by the same factor. But this size will be established only as a result of beam–beam interaction, and how can the beams be safely brought into collision? With the "initial" $\sigma_z$, $\xi_{x,y}$ will be far above the limits and the beams will be blown up and killed before they are stabilized by BS. To avoid this, the bunch population should be gradually increased *during collision*, so we come to *bootstrapping* [8].

Thus, in order not to lose luminosity, we have to work under conditions where the beams are lengthened several times due to BS. In turn, BS depends on the intensity and all three sizes of the counter bunch, $\sigma_{x,y}$ are affected by $\xi_{x,y}$ which strongly depend on $\sigma_z$, and $\sigma_y$ also depends on $\sigma_x$ due to betatron coupling. Another important point: if the beams have the same $\sigma_y$, then BS does not depend on $\sigma_y$, but if the vertical beam sizes are different, then BS will increase significantly for the inflated beam and decrease for the compressed one. All of this creates positive feedback that can lead to instability—3D flip-flop [8]. The main triggers are asymmetry in $N_p$ and $\sigma_y$ of colliding bunches, but scenarios may differ. One of them is shown in Fig. 2. Here, the bunches have a slight asymmetry in the population (about 5%). If the working point is chosen well, this does not lead to an increase in $\sigma_{x,y}$, although there will be some asymmetry in $\sigma_z$ (blue and magenta lines). But if the betatron tunes are slightly shifted (red and green lines) so that the footprint reaches some resonance, then $\sigma_y$ of the "weak" beam begins to increase, which enhances the BS and the $\sigma_z$ asymmetry. This process develops until $\xi_x$ grows enough to touch the synchrotron satellite of half-integer resonance, which causes a sharp increase in $\sigma_x$ for a weak beam. In the end, all three sizes of the weak beam grow many times over, and it dies. In other cases, $\sigma_x$ grows first, then $\sigma_y$. But synchrotron satellites of half-integer resonance always play a key role and, as we will see below, not only here.

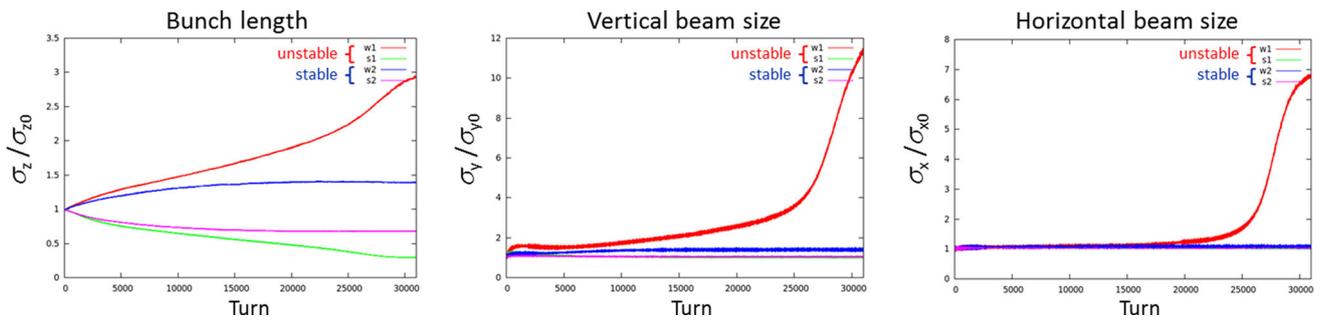

**Fig. 2** Example of 3D flip-flop: simulation results of a quasi-strong-strong model at low energy (45.6 GeV)

## 4 Parameter optimization

Another problem appears in collisions with LPA: coherent beam–beam instability, which is TMCI (Transverse mode coupling instability) caused by beam–beam interaction [9]. It develops in the horizontal plane and is manifested by wriggle of the bunch shape. The effect is 2D and $\varepsilon_x$ increases many times. Then, the betatron coupling leads to $\varepsilon_y$ increase in the same proportion, so the luminosity falls several times. This instability does not cause dipole oscillations and therefore cannot be mitigated by feedback. The only solution: find conditions under which it does not arise. Again, the threat comes from the synchro-betatron resonances $2(\nu_x - m \cdot \nu_z) = n$, which are very strong in collisions with LPA. On the other hand, in this case $\xi_x$ becomes very small, so our task is to place (with some margin) the footprint between adjacent resonances, the distance between which is equal to the synchrotron tune $\nu_z$. In other words, we need to decrease the $\xi_x/\nu_z$ ratio, and here we recall the expression for $\xi_x$ when $\phi \gg 1$ [5]:

$$\xi_x = \frac{r_e}{2\pi\gamma} \cdot \frac{N_p \beta_x^*}{\sigma_x^2(1+\phi^2)} \xrightarrow{\theta \ll 1, \phi \gg 1} \frac{2r_e}{\pi\gamma\theta^2} \cdot \frac{N_p \beta_x^*}{\sigma_z^2} \propto \xi_y \frac{\beta_x^*}{\sigma_z} \quad (5)$$

We do not want to decrease $\xi_y$, as this is the luminosity. It follows that it is necessary to decrease $\beta_x^*$, and this was done whenever possible [2, 335–336], but it was not enough. The next important step is to increase the momentum compaction factor $\alpha_p$. An advantage is that $\nu_z$ grows together (and by the same factor) with $\sigma_z$ and $1/\xi_x$. In addition, larger $\alpha_p$ increases the threshold of microwave instability to an acceptable level. And finally, lowering the RF voltage helps. This decreases $\nu_z$ and $\xi_x$ in the same proportion, but increases the order of resonances near the working point.

This can be understood from Fig. 3, where the simulation results for Z energy are shown. Note that all tunes here are for a super-period (from IP to IP). To have enough free space for the footprint (required for large $\xi_y$), $\nu_x$ should be less than $0.58 \div 0.59$ [2, 327–328] and we need to find a good working point here. As seen, this can only be done with a low $U_{RF}$. And it is very handy that this technique also helps to mitigate 3D flip-flop. It is curious that the width of good regions in Fig. 3 actually does not depend on $N_p$, compare blue and green lines. The reason is that in our parameter area $\sigma_z \propto \sqrt{N_p}$ because of BS, and therefore $\xi_x$ does not depend on $N_p$.

Optimization of parameters for each energy has its own features [2, 329–330]. As the energy increases, the bunch lengthening due to BS and $\phi$ decreases, and the damping decrements increase. For $t\bar{t}$, coherent beam–beam instability and 3D flip-flop are no longer dangerous, but another problem comes to the fore: limitation of the beam lifetime due to BS. And here it is required, on the contrary, to increase $\beta_{x,y}^*$ [8], and there is no longer an opportunity (and need) to vary the RF voltage. But in all cases BS requires minimizing asymmetry in $N_p$ and $\sigma_y$ for colliding bunches: at low energies this leads to 3D flip-flop, at high energies—to a decrease in the lifetime.

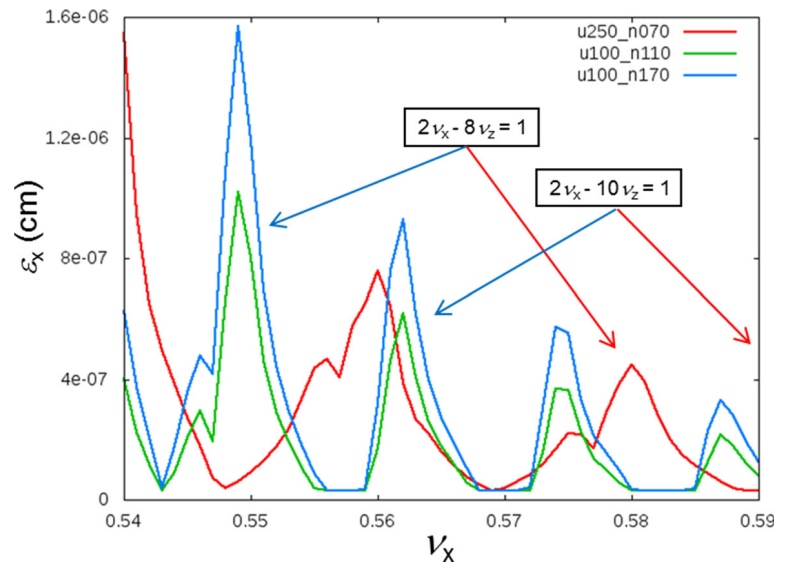

**Fig. 3** Growth of $\varepsilon_x$ due to coherent beam–beam instability, depending on $\nu_x$. $U_{RF} = 250$ MV (red) and 100 MV (green, blue). $N_p = 7 \cdot 10^{10}$ (red), $1.1 \cdot 10^{11}$ (green) and $1.7 \cdot 10^{11}$ (blue)

## 5 Open issues and next steps

If the collider has several IPs, then it is important to maintain symmetry so that the sections from IP to IP represent a super-period. The baseline FCC-ee design currently foresees two IPs. One of the obvious ways to increase the overall luminosity is to double their number, but certain problems arise here. First, it would mean a decrease in the synchrotron tune *per super-period* and intensified beamstrahlung: an increase in the energy spread and bunch length. As a result, even in the case of perfect 4-fold symmetry, the luminosity per IP decreases by 10–20%, depending on the energy. However, the main problem is that with four IPs, the super-periodicity requirements increase dramatically. This is due to the size of the full footprint, which will almost double and therefore can cross low-order resonances such as 1/2 and 1/3. The width of these resonances depends on the level of 4-fold symmetry breaking. Since the footprint is large, particles will survive, but the beams may swell and the luminosity will drop. One possible solutions is to shift the working point to avoid harmful resonances, but this leads to a decrease in the beam–beam limit. Certain questions should be considered before making an informed decision. What level of symmetry do we need for 4 IPs to give a noticeable increase in luminosity? What accuracy of the orbit and lattice correction can be achieved? At what cost? If four IPs would potentially allow higher luminosity, but the commissioning time is longer, will we get a higher integrated luminosity?

It should be noted that parameter optimization has so far been done for a simplified model. Dynamic aperture and momentum acceptance were of course taken into account, but when considering the beam–beam interaction, the lattice was assumed to be linear. This was fully justified for the first stage—the selection of the main parameters. But now we need to work out all the details, take into account more effects and their mutual influence. These include: misalignments and corrections (orbit, lattice, betatron coupling), chromatic coupling, impedance, synchrotron radiation in the quadrupoles, top-up injection with beam–beam interaction, etc. And the question of the number of IPs needs to be addressed in conjunction with all of the above.

Exploring all of these issues will require developing our modeling capabilities. Of course, the experience of DA$\Phi$NE, SuperKEKB, other colliders and light sources will be very useful, but for studying beam dynamics with BS, we cannot create prototypes and test facilities, because strong BS can only be observed in the FCC-ee when it is built. So for now, we can only rely on analytical estimates and modeling. We need different tools for different tasks, as well as several tools for the same task—to be able to cross-check the results.

**Acknowledgements** The author would like to thank K. Ohmi, K. Oide, F. Zimmermann and the entire FCC-ee team for many helpful discussions and fruitful collaboration. And special thanks to P. Charitos and F. Zimmermann for their comments and advice in preparing this essay.

**Funding** Open access funding provided by CERN (European Organization for Nuclear Research). This project is co-funded from the European Union's Horizon 2020 research and innovation programme under grant agreement No 951754.

**Data Availability Statement** This manuscript has associated data in a data repository. [Authors' comment: All data included in this manuscript are available upon request by contacting the corresponding author.]